\documentclass[
    ,final            % use final for the camera ready runs
  ]
  {aipproc}

\layoutstyle{8x11double}

\begin{document}

\title{Meta-stable Supersymmetry Breaking in an $\mathcal{N} = 1$ 
Perturbed Seiberg-Witten Theory}


\author{Shin Sasaki}{
  address={University of Helsinki and Helsinki Institute of Physics, P.O.Box 64, FIN-00014, Finland}
%  ,altaddress={<author1 address>} % additional visiting address
}

\author{Masato Arai}{
  address={CQUeST, Sogang University,
Shinsu-dong 1, Mapo-gu, Seoul 121-742, Korea}
}

\author{Claus Montonen}{
  address={University of Helsinki, P.O.Box 64, FIN-00014, Finland}
%  ,altaddress={<author1 address>} % additional visiting address
}

\author{Nobuchika Okada}{
  address={Theory Division, KEK, Tsukuba 305-0801, Japan}
%  ,altaddress={<author1 address>} % additional visiting address
}

\begin{abstract}
In this contribution, we discuss the possibility of meta-stable supersymmetry 
(SUSY) breaking vacua in a perturbed Seiberg-Witten theory with Fayet-Iliopoulos (FI) term.
We found meta-stable SUSY breaking vacua at the degenerated dyon and 
monopole singular points in the moduli space at the nonperturbative level.
\end{abstract}

\maketitle

%%%%%%%%%%%%%%%%%%%%%%%%%%%%%%%%%%%%%%%%%%%%
%% MAINMATTER
%%%%%%%%%%%%%%%%%%%%%%%%%%%%%%%%%%%%%%%%%%%%

\section{The model}
SUSY breaking at meta-stable vacua in various SQCD models has been intensively studied since 
the proposal of the ISS model \cite{InSeSh}. 
The other interest is meta-stable SUSY breaking in 
perturbed Seiberg-Witten theories \cite{OoOoPa, MaOoOoPa, Pa}.
In the following, we focus on this possibility.

We consider four-dimensional $\mathcal{N} = 2$ $SU(N_c) \times U(1)$, 
$N_f$ flavors SQCD with FI term. 
Supersymmetry in the model is partially broken down to $\mathcal{N} = 1$ 
due to the presence of adjoint mass terms. 
The extra $U(1)$ part is necessary for the FI term and treated as cut-off 
theory with Landau pole $\Lambda_L$. With the help of the Seiberg-Witten solution, we can analyze the theory in 
exact way provided the Landau pole is very far away and the perturbation 
terms are very much smaller than the $SU(N_c)$ dynamical scale $\Lambda$.
In the following, we focus on $N_c = N_f = 2$ case and show that there 
are SUSY breaking meta-stable minima in the full quantum level.
\subsection{$\mathcal{N} = 1$ SUSY preserving deformation of $\mathcal{N} 
= 2$ SQCD}
Let us consider a tree-level Lagrangian 
\begin{eqnarray}
\mathcal{L} = \mathcal{L}^{\mathcal{N} = 2}_{\mathrm{SQCD}} + \mathcal{L}_{\mathrm{soft}}.
\end{eqnarray}
Here $\mathcal{L}^{\mathcal{N} = 2}_{\mathrm{SQCD}}$ is the Lagrangian for $\mathcal{N} = 2$ $SU(2) \times 
U(1)$ super Yang-Mills with $N_f = 2$ massless fundamental hypermultiplets
\begin{eqnarray}
\mathcal{L}^{\mathcal{N} = 2}_{\mathrm{SQCD}} &=&  \frac{1}{2 \pi} \mathrm{Im}
\left[ \mathrm{Tr} \left\{ \tau_{22} \left(
\int \! d^4 \theta \ A_2^{\dagger} e^{2 V_2} A_2 e^{ - 2 V_2} 
\right. \right. \right. \nonumber \\
& & \left. \left. \left.
+ \frac{1}{2} \int \! d^2 \theta \ W_2^2 \right)
\right\}
\right] \nonumber \\
& & + \frac{1}{4 \pi} \mathrm{Im} \left[
\tau_{11} \left( \int \! d^4 \theta \ A_1^{\dagger} A_1
+ \frac{1}{2} \int \! d^2 \theta \ W_1^2 \right)
\right] \nonumber \\
& & +  \int \! d^4 \theta \left[
Q_r^{\dagger} e^{2 V_2 + 2 V_1} Q^r + \tilde{Q}_r e^{- 2 V_2 - 2 V_1}
\tilde{Q}^{r \dagger} \right] \nonumber \\
& & +  \sqrt{2} \left[ \int d^2 \theta \ \tilde{Q}_r (A_2 + A_1) Q^r + h.c.
\right],
\end{eqnarray}
where $V_2, A_2$ and $V_1, A_1$ are vector and chiral superfields belonging to the 
$SU(2)$ and $U(1)$ vector multiplets respectively.
The chiral superfields $Q^r_I$ and $\tilde{Q}_r^I$
 are hypermultiplets that are in the fundamental and anti-fundamental
 representations of the $SU(2)$ gauge group ($r=1,2$ is the flavor index,
 and $I=1,2$ is the $SU(2)$ color index). $W$ is the $\mathcal{N} = 1$ superfield 
 strength and $\tau_{ij}$ are complex gauge couplings.

The second term $\mathcal{L}_{\mathrm{soft}}$ is the soft SUSY breaking term given by
\begin{eqnarray}
\mathcal{L}_{\mathrm{soft}} = \int \! d^2 \theta \left(
\mu_2 \mathrm{Tr} (A_2^2) + \frac{1}{2} \mu_1 A_1^2 + \lambda A_1
\right) + h.c.
\end{eqnarray}
In  $\mathcal{L}_{\mathrm{soft}}$, $\mu_1, \mu_2$ are masses 
corresponding to $U(1)$ and $SU(2)$ part of the adjoint scalars and $\lambda$ is the FI parameter.
In the absence of $\mathcal{L}_{\mathrm{soft}}$, the gauge symmetry is 
broken as $SU(2) \times U(1) \to U(1)_c \times U(1)$ on the 
Coulomb branch
\begin{eqnarray}
q_r = \tilde{q}_r = 0, \quad 
A_2 = \left(
\begin{array}{cc}
a_2 & 0 \\
0 & - a_2
\end{array}
\right), \quad A_1 = a_1,
\end{eqnarray}
where $q, \tilde{q}$ are hypermultiplet scalars.
Once we turn on $\mathcal{L}_{\mathrm{soft}}$, 
there are SUSY vacua on the Coulomb and Higgs branches.
We are going to investigate the quantum effective action 
on the Coulomb branch. 

\section{Quantum Theory}
The exact low energy effective Lagrangian is described by light 
 fields, the $SU(2)$ dynamical scale $\Lambda$, the Landau pole 
 $\Lambda_L$, the masses $\mu_i \ (i=1,2)$ and 
 the FI parameter $\lambda$.
If the perturbation terms are much smaller than the dynamical scale 
$\Lambda$, the effective Lagrangian ${\cal L}_{\mathrm{exact}}$ is given 
by 
\begin{eqnarray}
 {\cal L}_{\mathrm{exact}}={\cal L}_{\mathrm{SUSY}}+{\cal L}_{\mathrm{pert.}}
 +{\cal O}(\mu_i^2,\lambda).
\end{eqnarray}
Here the first term ${\cal L}_{\mathrm{SUSY}}$ describes an ${\cal N}=2$
 SUSY Lagrangian containing full quantum corrections.
The second term ${\cal L}_{\rm pert.}$ includes the masses and the FI
 terms in the leading order.

First we consider the general formulas for the effective Lagrangian
 ${\cal L}_{\mathrm{SUSY}}$.
The Lagrangian ${\cal L}_{\mathrm{SUSY}}$ is given by two parts,
 vector multiplet part $\mathcal{L}_{\mathrm{VM}}$ and
 hypermultiplet part $\mathcal{L}_{\mathrm{HM}}$,
\begin{eqnarray}
\mathcal{L}_{\mathrm{SUSY}} = \mathcal{L}_{\mathrm{VM}} + \mathcal{L}_{\mathrm{HM}}.
\end{eqnarray}
The ${\cal L}_{\mathrm{VM}}$ part consists of $U(1)_c$ and $U(1)$ vector multiplets.
The effective Lagrangian for these vector multiplets is
\begin{eqnarray}
\mathcal{L}_{\mathrm{VM}} &=& \frac{1}{4 \pi} \mathrm{Im} \sum_{i,j=1}^2
\left[ \int \! d^4 \theta \ \frac{\partial \mathcal{F}}{\partial A_i} A_i^{\dagger}
\right. \nonumber \\
& & \left. + \frac{1}{2} \int \! d^2 \theta \ \tau_{ij} W_i W_j \right], \label{VM}
\end{eqnarray}
where $\mathcal{F} = \mathcal{F} (A_2, A_1, \Lambda, \Lambda_L)$ is a prepotential
 as will be discussed below. 
The effective gauge coupling $\tau_{ij}$ is defined by $\tau_{ij} = 
\frac{\partial^2 \mathcal{F}}{\partial a_i \partial a_j}$ with moduli $a_i$.
The hypermultiplet part $\mathcal{L}_{\mathrm{HM}}$ is
\begin{eqnarray}
\mathcal{L}_{\mathrm{HM}} &=& \int \! d^4 \theta \left[
M^{\dagger}_r e^{2 n_m V_{2D} + 2 n_e V_2 + 2 n V_1} M^r
\right. \nonumber \\
& & \left. 
+ \tilde{M}_r e^{- 2 n_m V_{2D} - 2 n_e V_2 - 2 n V_1} \tilde{M}^{r \dagger}
\right] \nonumber \\
& &  + \sqrt{2} \int \! d^2 \theta \left[ \tilde{M}_r (n_m A_{2D} + n_e A_2 + n A_1) M^r \right.
\nonumber \\
& & \left. + h.c. \right],
\end{eqnarray}
where $M^r, \tilde{M}_r$ are chiral superfields and
 $V_{2D}, A_{2D}$ are dual variables of $V_2, A_2$. 
These hypermultiplets correspond to the light BPS dyons, monopoles and
 quarks.
 which are specified through the appropriate quantum numbers
 $(n_e, n_m)_n$.
Here $n_e$ and $n_m$ are the electric and magnetic
 charges of $U(1)_c$, respectively, whereas  $n$ is the $U(1)$ charge.
The potential is a function of $M, \tilde{M}, a_1, a_2$.
We found stationary points along $M, \tilde{M}$ directions at (1) $M = 
\tilde{M} = 0$ and (2) $M \not= 0, \tilde{M} \not= 0$.
The potential value at each stationary points are evaluated as
\begin{eqnarray}
& & {(1)}~V(a_2,a_1)=U, \\
& & {(2)}~V(a_2,a_1)=U-4S \mathcal{M}^4, \label{potential2}
\end{eqnarray}
where $U = U(a_1,a_2), S = S(a_1, a_2) > 0$ are functions of $a_1, a_2$ and $\mathcal{M} 
\equiv |M| = |\tilde{M}|$ \cite{ArMoOkSa2}.
The stationary point (\ref{potential2}) where the light hypermultiplet acquires
 a vacuum expectation value by the condensation of the BPS states is energetically favored because $S>0$.

Due to the abovementioned reason, we focus on the singularity points in 
the moduli space. 
To find the explicit potential, we need the moduli space metric, and 
hence the prepotential. The monodromy transformation around the singular 
points in the moduli space dictates us that the $U(1)$ modulus $a_1$ can be interpreted as 
the common hypermultiplet mass $m$ in the $SU(2)$ gauge theory. This fact 
implies that the prepotential in our model is given by 
\begin{eqnarray}
\mathcal{F} (a_2,a_1,\Lambda, \Lambda_L)
 = \mathcal{F}_{SU(2)}^{(SW)} (a_2, m,\Lambda)
    \Bigg{|}_{m=\sqrt{2} a_1}+C a_1^2,
\end{eqnarray}
where $\mathcal{F}_{SU(2)}^{(SW)}$ is the prepotential for $SU(2)$ 
massive SQCD with common mass $m \equiv \sqrt{2} a_1$. The constant $C$ 
is a free parameter which is used to fix the Landau pole $\Lambda_L$ to 
the appropriate value\footnote{We fix $C = 4 \pi i$ which implies $\Lambda_L \sim 10^{17-18} \Lambda$.}. 

The singular points on the moduli space are determined
 by a cubic polynomial \cite{SeWi}.
The solutions of the cubic polynomial give the positions of the singular
 points in the $u$-plane.
In the case $N_c = N_f=2$ with a common hypermultiplet mass $m$,
 which is regarded as the modulus $\sqrt{2} a_1$ here,
 the solution is obtained as
\begin{eqnarray}
u_1 = - m \Lambda - \frac{\Lambda^2}{8}, \
u_2 = m \Lambda - \frac{\Lambda^2}{8}, \
u_3 = m^2 + \frac{\Lambda^2}{8}.
\end{eqnarray}
The singular points correspond to dyons, a monopole and a quark. The 
behavior of the singularity flow along $a_1$ direction can be found in \cite{ArMoOkSa2}.
At the singular points in the moduli space, $a_2$ and $a_1$ are related to
each other and the potential is a function of $a_1$ only. 
To find the stationary points of the potential along the $a_1$ direction 
is a difficult task and we need the help of numerical analysis.

\begin{figure}[htb]
\includegraphics[scale=.6]{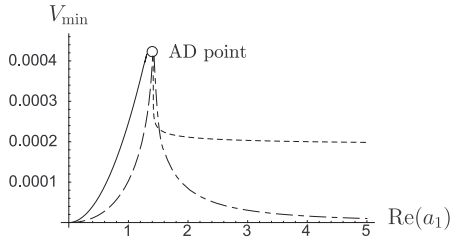}
\caption{Global structure of vacuum.
Solid and dashed curves show the evolutions of the potential energies at the
 monopole and left(right) dyon points for $0\le {\rm Re}(a_1)<\Lambda/(2\sqrt{2})$.
The potential energies at the right dyon(dotted)
 and quark(dash-dotted) points for ${\rm Re}(a_1)>\Lambda/(2\sqrt{2})$ are also plotted.
We have fixed $\Lambda = 2 \sqrt{2}$.}
\label{global_structure_potential}
\end{figure}

Let us start from the $\mu_1 = \lambda = 0$ case.
Fig. \ref{global_structure_potential} shows the global structure of the 
potential along the $\mathrm{Re}(a_1)$ direction. 
%%%%%%%%%%%%%%%%%%%%%%%%%%%%%%%%%%%%%%%%%%%%%%%%%%%%%
%%%%%%%%%%%%%%%%%%%%%%%%%%%%%%%%%%%%%%%%%%%%%%%%%%%%%
As a result, we found the global SUSY minima at $a_1 = 0$ in the 
degenerated dyon and monopole singular points. 
%Turning on the imaginary direction does not change the result.

\begin{figure}[htb]
\includegraphics[scale=.45]{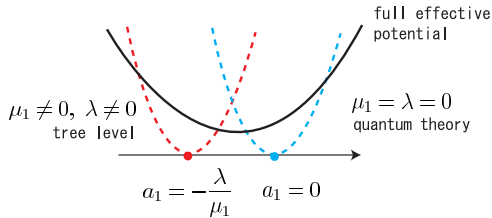}
\caption{Schematic picture of SUSY breaking mechanism}
\label{IYIT}
\end{figure}

Next, let us turn on $\mu_1$ and $\lambda$. 
In the presence of the soft term,
 the gauge dynamics favors the monopole and the dyon points at $a_1=0$
 as SUSY vacua besides the runaway vacua.
It implies that if we add $\mu_1 \not=0, \lambda \not= 0$ terms 
 which produce a vacuum at a point different from $a_1=0$
 at the classical level, SUSY is dynamically broken as a consequence of
 the discrepancy of SUSY conditions between the classical and the
 quantum theories.
Actually, turning on the mass $\mu_1$ and the FI parameter
 $\lambda$ realizes such a situation.
In this case, the classical vacuum is at $a_1=-\lambda/\mu_1$,
 different from the point $a_1=0$ which the dynamics favors.
A resultant SUSY breaking vacuum is realized at
 non-zero value of $a_1$.
This is
 very similar to the SUSY breaking mechanism discussed in the
 Izawa-Yanagida-Intriligator-Thomas model in ${\cal N}=1$ SUSY gauge
 theory \cite{IzYa, InTh}.
We show a schematic picture of our situation in Fig. \ref{IYIT}.
%%%%%%%%%%%%%%%%%%%%%%%%%%%%%%%%%%%%%%%%%%%%%%%%%%%%%%%%%%%%%%%%%%%%%%
%%%%%%%%%%%%%%%%%%%%%%%%%%%%%%%%%%%%%%%%%%%%%%%%%%%%%%%%%%%%%%%%%%%%%%

%%%%%%%%%%%%%%%%%%%%%%%%%%%%%%%%%%%%%%%%%%%%%%%%%%%%%%%%%%%%%%%%%%%%%%
\begin{figure}[htb]
\includegraphics[scale=.5]{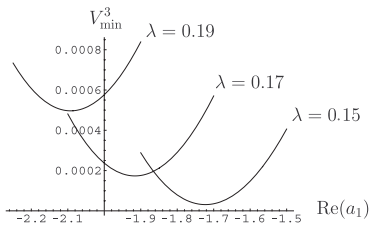}
\caption{Local SUSY breaking minimum at the monopole singular point for
 $\mu_1=\mu_2=0.1$ and $\lambda=0.15,0.17,0.19$ from bottom to top. Here 
 $0 \le \mathrm{Re} (a_1) < \Lambda/2 \sqrt{2}$.}
\label{monopole_minimum}
\end{figure}
%%%%%%%%%%%%%%%%%%%%%%%%%%%%%%%%%%%%%%%%%%%%%%%%%%%%%%%%%%%%%%%%%%%%%%

\begin{figure}[htb]
\includegraphics[scale=.35]{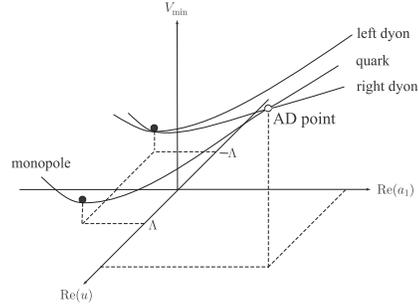}
\caption{Qualitative picture of the evolutions of the potential minima.}
\label{3d-plot}
\end{figure}

Let us see in detail how this works for non-zero values of $\mu_1, \mu_2$ and $\lambda$.
Fig. \ref{monopole_minimum} shows the evolution of the potential energies
 at the monopole point $V_{\mathrm{min}}^3$ for several values of
 $\lambda$ as a function of ${\rm Re}(a_1)$ with $\mu_1=\mu_2=0.1$.
The potential minimum is no longer realized at $a_1=0$, but the location
 is shifted to negative values of ${\rm Re}(a_1)$ as is expected from
 the discussion in the previous paragraph.
Furthermore, the potential energy has a non-zero value and therefore
 SUSY is dynamically broken. We find that the potential energies at the 
 left and right dyon singular points also have the same structure.
A qualitative picture of the evolutions of the potential minima
 is depicted in Fig. \ref{3d-plot}.

%%%%%%%%%%%%%%%%%%%%%%%%%%%%%%%%%%%%%%%%%%%%%%%%%%%%%%%%%%%%%%%%%%%%%%
%%%%%%%%%%%%%%%%%%%%%%%%%%%%%%%%%%%%%%%%%%%%%%%%%%%%%%%%%%%%%%%%%%%%%%

In addition to these local minima, there are supersymmetric vacua on the 
Higgs branch which survive from the quantum corrections. We estimated the 
decay rate from our local minima to the SUSY vacua on the Higgs branch 
and found that the decay rate can be taken to be very small. This means our 
local minima are nothing but meta-stable SUSY breaking vacua. 

%%%%%%%%%%%%%%%%%%%%%%%%%%%%%%%%%%%%%%%%%%%%%%%%
%% BACKMATTER
%%%%%%%%%%%%%%%%%%%%%%%%%%%%%%%%%%%%%%%%%%%%%%%%
%
%
\begin{theacknowledgments}
The work of S.~S is supported by the bilateral program of Japan Society
for the Promotion of Science and Academy of Finland, ``Scientist
Exchanges''.
\end{theacknowledgments}

\begin{thebibliography}{0}
\bibitem{InSeSh}
  K.~Intriligator, N.~Seiberg and D.~Shih,
  %``Dynamical SUSY breaking in meta-stable vacua,''
  JHEP {\bf 0604} (2006) 021
  [arXiv:hep-th/0602239].
%
\bibitem{OoOoPa}
  H.~Ooguri, Y.~Ookouchi and C.~S.~Park,
  %``Metastable Vacua in Perturbed Seiberg-Witten Theories,''
  arXiv:0704.3613 [hep-th].
  %%CITATION = ARXIV:0704.3613;%%
%
\bibitem{MaOoOoPa}
  J.~Marsano, H.~Ooguri, Y.~Ookouchi and C.~S.~Park,
  %``Metastable Vacua in Perturbed Seiberg-Witten Theories, Part 2:
  %Fayet-Iliopoulos Terms and K\'ahler Normal Coordinates,''
  Nucl.\ Phys.\  B {\bf 798} (2008) 17
  [arXiv:0712.3305 [hep-th]].
  %%CITATION = NUPHA,B798,17;%%
%
\bibitem{Pa}
  G.~Pastras,
  %``Non supersymmetric metastable vacua in N = 2 SYM softly broken to N = 1,''
  arXiv:0705.0505 [hep-th].
  %%CITATION = ARXIV:0705.0505;%%
%
\bibitem{ArMoOkSa}
  M.~Arai, C.~Montonen, N.~Okada and S.~Sasaki,
  %``Meta-stable Vacuum in Spontaneously Broken N=2 Supersymmetric Gauge
  %Theory,''
  Phys.\ Rev.\  D {\bf 76} (2007) 125009
  [arXiv:0708.0668 [hep-th]].
%
\bibitem{ArMoOkSa2}
  M.~Arai, C.~Montonen, N.~Okada and S.~Sasaki,
  %``Dynamical Supersymmetry Breaking from Meta-stable Vacua in an N=1
  %Supersymmetric Gauge Theory,''
  JHEP {\bf 0803} (2008) 004
  [arXiv:0712.4252 [hep-th]].
  %%CITATION = JHEPA,0803,004;%%
%
\bibitem{SeWi}
  N.~Seiberg and E.~Witten,
 %``Monopoles, duality and chiral symmetry breaking in N=2 supersymmetric
 % QCD,''
  Nucl.\ Phys.\ B {\bf 431} (1994) 484
  [arXiv:hep-th/9408099].
%
\bibitem{IzYa}
  K.~I.~Izawa and T.~Yanagida,
  %``Dynamical Supersymmetry Breaking in Vector-like Gauge Theories,''
  Prog.\ Theor.\ Phys.\  {\bf 95} (1996) 829
  [arXiv:hep-th/9602180].
%
\bibitem{InTh}
  K.~A.~Intriligator and S.~D.~Thomas,
  %``Dynamical Supersymmetry Breaking on Quantum Moduli Spaces,''
  Nucl.\ Phys.\  B {\bf 473} (1996) 121
  [arXiv:hep-th/9603158].

\end{thebibliography}
\end{document}